\documentclass[11pt,times]{emulateapj}

\usepackage{natbib}
\usepackage{epsfig}

 \newcommand{\be}{\begin{equation}}
 \newcommand{\ee}{\end{equation}}
 
 \newcommand{\eg}{\emph{e.g.}}
 \newcommand{\kms}{\mbox{km\ \ensuremath{\rm{s}^{-1}}}}
 
\shortauthors{Cordiner et al.}

\begin{document}

\title{Discovery of interstellar anions in Cepheus star-forming region}

\author{M. A. Cordiner and S. B. Charnley}
\affil{Astrochemistry Laboratory and The Goddard Center for Astrobiology, NASA Goddard Space Flight Center, Code 691,
8800 Greenbelt Road, Greenbelt, MD 20771, USA}
\email{martin.cordiner@nasa.gov}

\author{J. V. Buckle}

\affil{Cavendish Astrophysics Group and Kavli Institute for Cosmology,
Institute of Astronomy, University of Cambridge, Madingley Road, 
Cambridge, CB3 0HE, UK}

\author{C. Walsh and T. J. Millar}
\affil{Astrophysics Research Centre, School of Mathematics and Physics, Queen's University Belfast, Belfast, BT7 1NN, Northern Ireland}

\begin{abstract}

We report the detection of microwave emission lines from the hydrocarbon anion C$_6$H$^-$ and its parent neutral C$_6$H in the star-forming region L1251A (in Cepheus), and the pre-stellar core L1512 (in Auriga). The carbon-chain-bearing species C$_4$H, HC$_3$N, HC$_5$N, HC$_7$N and C$_3$S are also detected in large abundances. The observations of L1251A constitute the first detections of anions and long-chain polyynes and cyanopolyynes (with more than 5 carbon atoms) in the Cepheus Flare star-forming region, and the first detection of anions in the vicinity of a protostar outside of the Taurus molecular cloud complex, highlighting a wider importance for anions in the chemistry of star formation.  Rotational excitation temperatures have been derived from the HC$_3$N hyperfine structure lines, and are found to be 6.2~K for L1251A and 8.7~K for L1512. The anion-to-neutral ratios are 3.6\% and 4.1\%, respectively, which are within the range of values previously observed in the interstellar medium, and suggest a relative uniformity in the processes governing anion abundances in different dense interstellar clouds. This research contributes towards the growing body of evidence that carbon chain anions are relatively abundant in interstellar clouds throughout the Galaxy, but especially in the regions of relatively high density and high depletion surrounding pre-stellar cores and young, embedded protostars.

\end{abstract}

\keywords{astrochemistry --- ISM: abundances --- ISM: molecules --- ISM: clouds --- stars: formation}

\section{Introduction}

Hydrocarbon anions have recently been discovered in the quiescent molecular clouds TMC-1 \citep{mcc06} and Lupus-1A \citep{sak10}, the carbon-rich AGB star IRC+10216 \citep{cer07,rem07}, the protostars L1527 and L1521F and the pre-stellar core L1544 \citep{sak08a,gup09}. The possibility that an appreciable fraction of molecular material in interstellar clouds might be in the form of anions was first suggested by \citet{sar80} and \citet{her81}, who pointed out that carbon chain molecules and other radicals have large electron affinities, leading to large radiative attachment rates such as those measured by \cite{wood80}. However, the full significance of anions for astrochemistry is still far from understood, so we seek to address the question of just how widespread anions are in the Galaxy. Apart from the single detection in Lupus \citep{sak10}, all interstellar anion detections to-date have been confined to the Taurus molecular cloud complex. The origin of the large observed abundances of polyynes (C$_n$H, $n=2-8$) and cyanopolyynes (HC$_{2n+1}$N, $n=1-5$) in this region \citep[\eg][]{cer86,bel98,bru07} is debated, and may be attributable to a young gas-phase chemistry \citep{her89}, interactions between gas and dust \citep[\eg][]{bro91}, or shocking in cloud-cloud collisions \citep{lit78}. \citet{wal09} theorised that the observed hydrocarbon anions act as catalysts for the synthesis of increased polyyne and cyanopolyyne abundances.  New observations of long carbon chains and anions outside of the Taurus complex will provide tests for these astrochemical models and will provide data for the development of new theories regarding the formation of carbon chains throughout the Galaxy.

Models for anion chemistry are able to reproduce, with reasonable accuracy, the observed abundances of C$_6$H$^-$ and C$_8$H$^-$ in TMC-1, IRC+10216 and L1527  \citep{mil07,rem07,har08,cor08}. The recent detection of CN$^-$ in IRC+10216 by \citet{agu10} also matches well the abundance predicted by the chemical model of \citet{cor09}. However, there are still discrepancies between the modelled and observed anion-to-neutral ratios, especially for C$_4$H$^-$ \citep{her08}, and the lack of anions in PDRs \citep{agu08} is at variance with the model predictions of \citet{mil07}. Clearly, our understanding of molecular anion chemistry is incomplete. Anions may be of wider importance in astrophysics; their presence within magnetised, collapsing cores may have consequences for star formation dynamics, through their influence on the ambipolar diffusion rate. Molecular anions, including CH$_2$CN$^-$, have been considered as plausible carriers for at least some of the unidentified diffuse interstellar bands \citep{cor07}. Anion abundances are theorised to be sensitive to electron attachment and photodetachment rates (see \citealt{mil07}), and may therefore provide a useful tool for the determination of accurate electron densities and cosmic ray/photo- ionisation rates in astrophysical environments.

Presently, many of the key reactions and rates relevant to molecular anion chemistry, such as those of radiative electron attachment, have not been well-studied in the laboratory, and so are poorly constrained or grossly approximated in chemical models. Given the current lack of laboratory experiments and detailed (quantum) theoretical calculations, the only way to constrain these parameters and further our understanding of the role of anions in astrophysics is by radio observations and complementary chemical modelling of molecular anions in various astrophysical environments. 

This Letter reports new detections of the carbon chain anion C$_6$H$^-$ in the vicinity of a young, low-mass protostar in the dense molecular cloud L1251A (in the Cepheus complex), and in the prestellar core L1512 (in the Taurus-Auriga complex).

\section{Observations}
\label{obs}

\begin{figure}
\centering
\includegraphics[angle=270, width=\columnwidth]{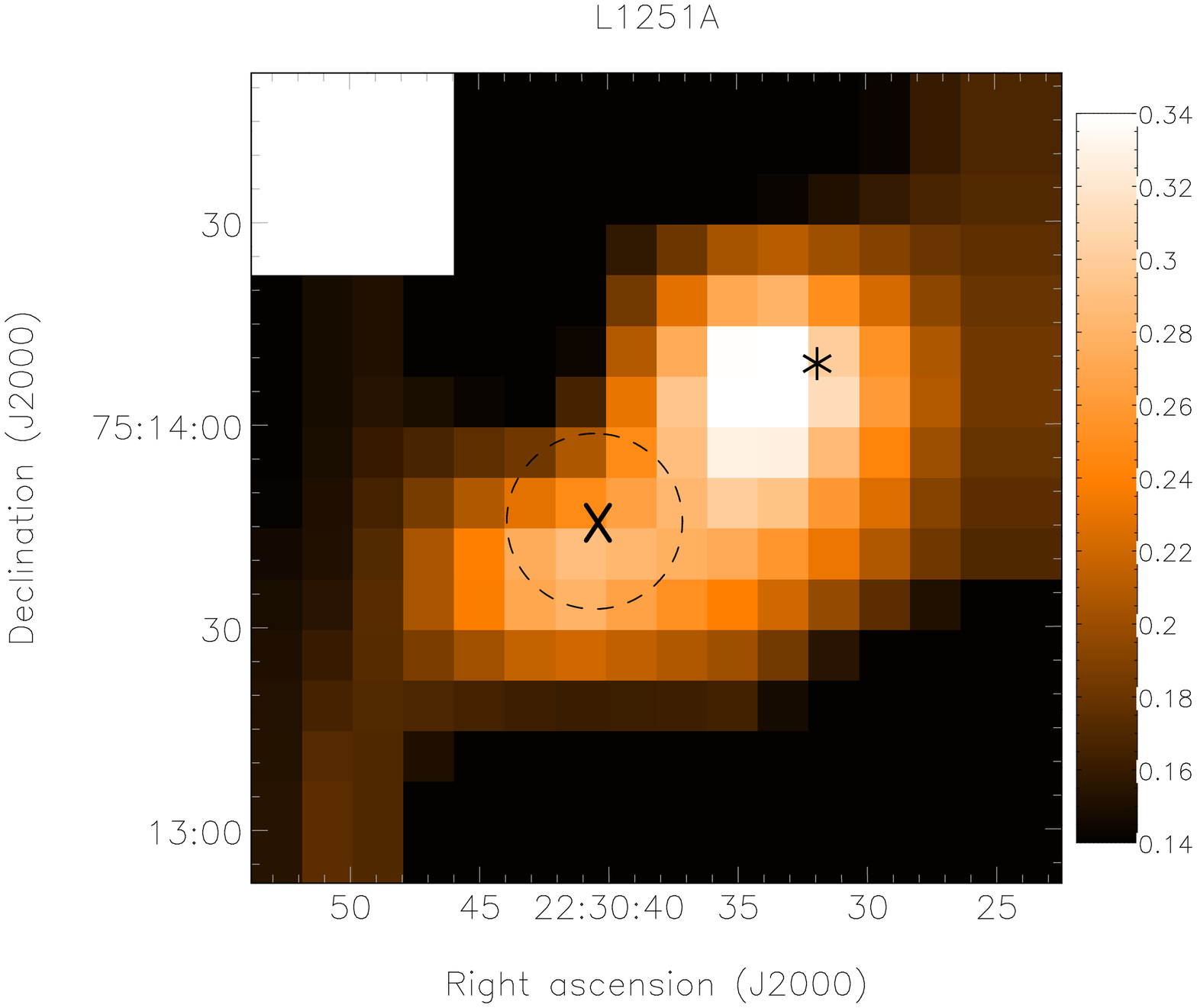}
\includegraphics[angle=270, width=\columnwidth]{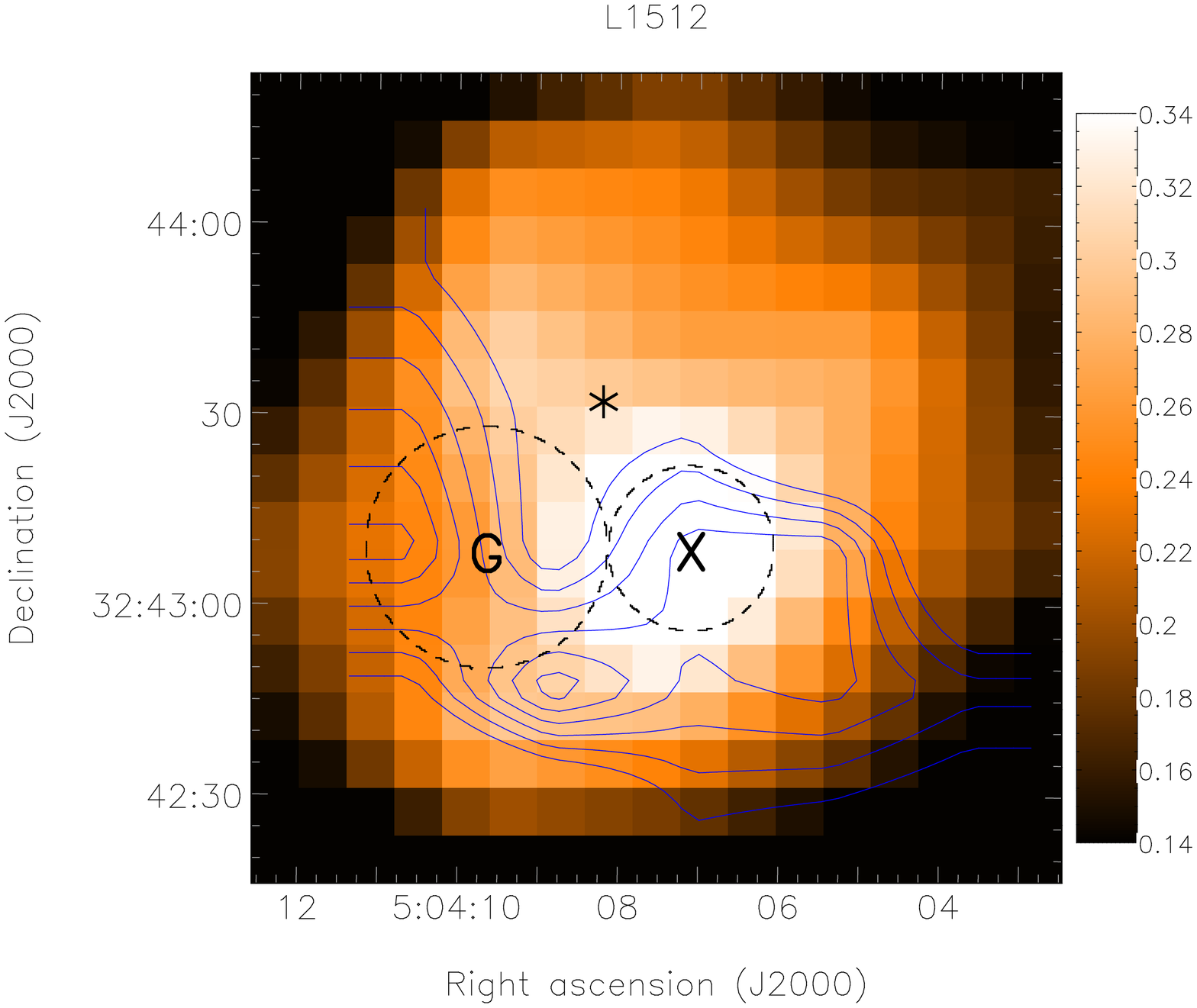}
\caption{Onsala HC$_3$N $J=10-9$ maps of L1251A (top) and L1512 with C$_4$H $N=9-8$ contours overlaid (bottom). Scale bar units are K\,\kms. C$_4$H contours show peak antenna temperature, starting at 0.36~K, and increasing in steps of 0.02~K.  The observed GBT positions are labeled `X' and location of the \citet{gup09} L1512 anion search is labeled `G'. Asterisks show positions of protostar (for L1251A) and sub-mm source (for L1512). Dashed circles represent the GBT beam FWHM of the respective anion searches. Our chosen L1251A position does not coincide precisely with the HC$_3$N peak because it was based on an earlier, lower-sensitivity map than that shown. \label{fig:map}}
\end{figure}

Sources were selected from a set of sixteen dense clouds for which HC$_3$N $J=10-9$ maps had previously been obtained using the Onsala 20-m telescope between 2005 and 2007 (some of which were published by \citealt{buc06}). The close chemical relationship between polyynes and cyanopolynes (see \emph{e.g.} \citealt{fed90,mil94}), suggests that C$_4$H and C$_6$H should be abundant in dense molecular clouds, close to where the HC$_3$N peaks. Therefore, to improve the chances of successfully detecting C$_4$H, C$_6$H and C$_6$H$^-$ compared with previous surveys \citep[\emph{e.g.}][]{gup09}, we targeted the strongest HC$_3$N emission peaks in these maps, which, due to chemical differentiation \citep[\emph{e.g.}][]{buc06,ber07}, are not generally coincident with the locations of peak emission from commonly observed molecules such as NH$_3$, N$_2$H$^+$ or CO (for which maps already exist in the literature). Onsala HC$_3$N $J=10-9$ maps of L1251A and L1512 are shown in Figure \ref{fig:map}. The adopted coordinates for our anion and carbon chain searches were: L1251A: RA(2000) = 22:30:40.4, DEC(2000) = +75:13:46; L1512:  RA(2000) = 5:04:07.1, DEC(2000) = 32:43:09. The L1251A position lies $40''$ from the centre of the Class 0 protostar L1251A IRS3, and the L1512 position lies $26''$ from the centre of a ($\approx120''$-wide) pre-stellar core \citep{dif08,kir05}. The positions observed by \citet{gup09} (who did not detect C$_6$H$^-$ in either cloud), are not coincident with our observed positions. 

Observations were carried out in the months of April and June 2010 using the NRAO 100-meter Green Bank Telescope\footnote{The National Radio Astronomy Observatory is a facility of the National Science Foundation operated under cooperative agreement by Associated Universities, Inc.}. The Ka receiver was used with 50 MHz bandwidth and 8192 channels (corresponding to a channel spacing of $\approx0.065$ \kms), in each of four spectral windows. In the middle of the observed frequency range (28 GHz), the telescope beam FWHM was $26''$ and the beam efficiency factor was 0.88. We performed deep integrations to search for emission lines from C$_4$H, C$_6$H, C$_6$H$^-$ and HC$_7$N. Additional observations were obtained of HC$_5$N, C$_3$S and CH$_3$CCH. For the compact, spatially isolated source L1512, beam switching (with $78''$ throw) was used, and for the more extended source L1215A, frequency switching was used. Pointing was checked every one to two hours and was typically accurate to within $5''$. Total system temperatures were in the range 40 to 60~K. 

New maps of HC$_3$N $J=10-9$ emission in L1251A and C$_4$H $N=9-8$ emission in L1512 were obtained using the Onsala 20~m telescope in May 2010 (shown in Figure \ref{fig:map}).  The basic observation and reduction techniques were presented by \citet{buc06}.

\section{Results}
\label{results}

\begin{figure}
\includegraphics[width=\columnwidth]{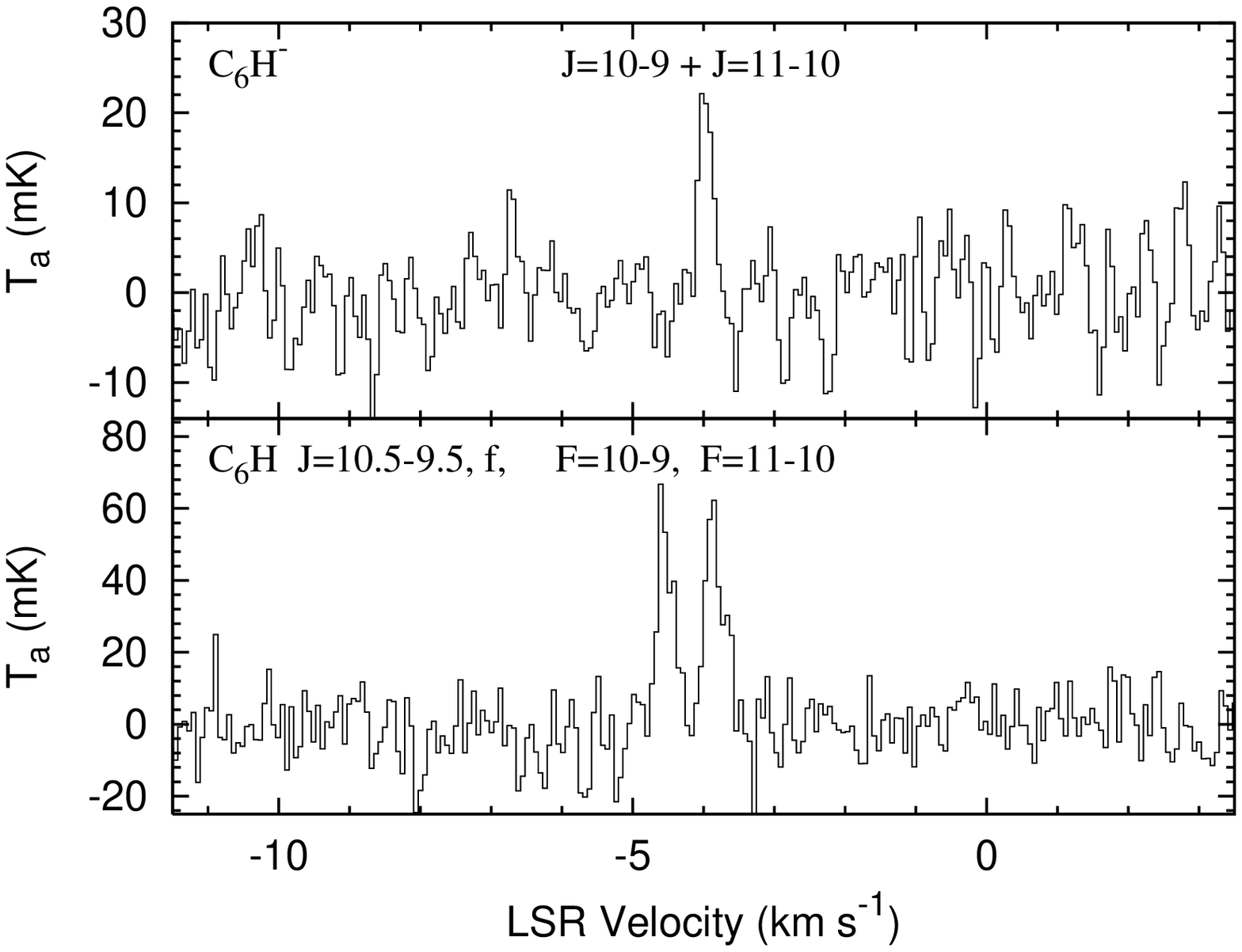}
\includegraphics[width=\columnwidth]{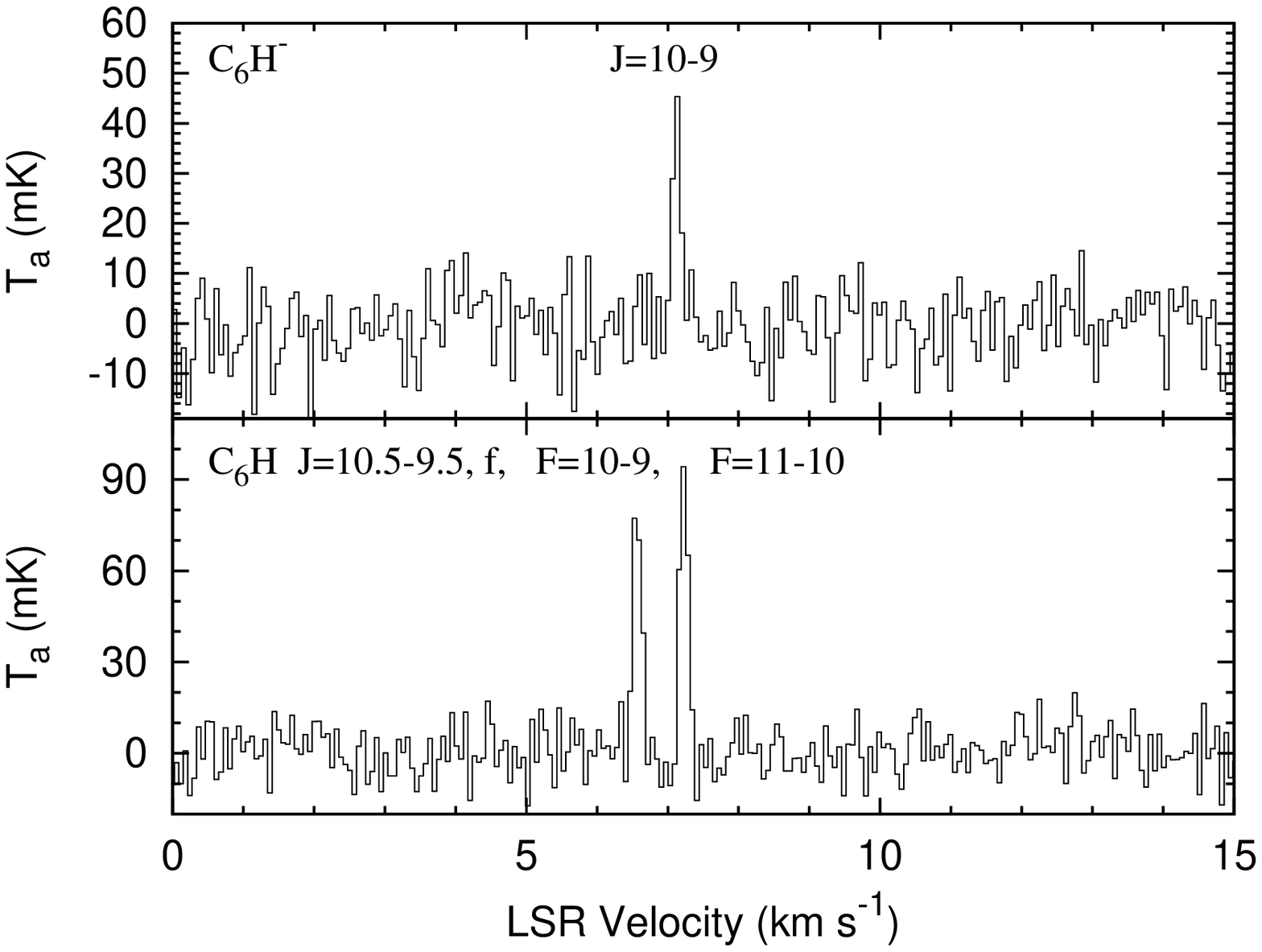}
\caption{Plot showing (averaged) C$_6$H$^-$ $J=10-9$ and $J=11-10$ spectrum of L1251A (top) and $J=10-9$ spectrum of L1512 (bottom). The C$_6$H $J=10.5-9.5,\,f$ spectra are also shown. \label{fig:spectra}}
\end{figure}

\begin{deluxetable*}{lllcllcllll}
\tabletypesize{\scriptsize}
\tablecaption{Observed line parameters\label{tab:lines}}
\tablewidth{0pt}
\tablehead{
&&\multicolumn{4}{c}{L1512}&&\multicolumn{4}{c}{L1251A}\\
\cline{3-6}\cline{8-11}
\colhead{Species \hfill Transition\tablenotemark{a}}& \colhead{Frequency} & \colhead{$T_a$} & \colhead{$v$} & \colhead{$\Delta v$} & \colhead{$\int T_adv$} & & \colhead{$T_a$} & \colhead{$v$} & \colhead{$\Delta v$} & \colhead{$\int T_adv$}
}
\startdata
C$_6$H$^-$ \hfill $10-9$ & 27537.13 & 46 (7) &   7.09 & 0.13 (2) & 6 (1) & &27 (7) &   $-$3.96 & 0.21 (6) & 6 (2) \\      
C$_6$H$^-$ \hfill $11-10$ & 30290.81 & 18 (7) &   7.10 & 0.16 (7) & 3 (1) & &24 (7) &   $-$4.00 & 0.18 (5) & 5 (1) \\      
C$_4$H\ \ \hfill $3-2,3.5-2.5,3-2$&  28532.31 & 715 (50) & 7.05 & 0.15 (1) & 114 (8) & & 383 (45) & $-$4.05 & 0.38 (5) & 155 (18) \\    
C$_4$H\ \ \hfill $3-2,3.5-2.5,4-3$&  28532.46 & 931 (50) & 7.06 & 0.15 (1) & 149 (8) & & 486 (34) & $-$4.04 & 0.41 (3) & 212 (15) \\   
C$_6$H \hfill $10.5-9.5,f,11-10$ &  29109.64 & 91 (7) &   7.26 & 0.13 (1) & 13 (1) & &  56 (6) &   $-$3.85 & 0.32 (4) & 19 (2) \\   
C$_6$H \hfill $10.5-9.5,f,10-9$  &  29109.69 & 84 (6) &   7.01 & 0.16 (1) & 14 (1) & &  58 (6) &   $-$4.11 & 0.27 (3) & 17 (2) \\     
C$_6$H \hfill $10.5-9.5,e,11-10$ &  29112.71 & 99 (7) &   7.19 & 0.14 (1) & 15 (1) & &  58 (6) &   $-$3.94 & 0.28 (3) & 17 (2) \\     
C$_6$H \hfill $10.5-9.5,e,10-9$  &  29112.75 & 84 (7) &   6.96 & 0.15 (1) & 13 (1) & &  52 (6) &   $-$4.15 & 0.33 (4) & 18 (2) \\     
HC$_3$N \hfill $3-2,3-3$ & 27292.90 & 671 (65) & 7.07 & 0.12 (1) & 86 (11) & & 246 (24) & $-$3.99 & 0.31 (3) & 81 (11) \\
HC$_3$N \hfill $3-2,2-1$ & 27294.07 & 2102 (52) & 6.98 & 0.16 (0) & 358 (9) & & 967 (22) & $-$4.10 & 0.37 (1) & 382 (10) \\   
HC$_3$N \hfill $3-2,3-2$ & 27294.29 & 2769 (53) & 7.07 & 0.16 (0) & 472 (9) & & 1331 (23) & $-$4.00 & 0.34 (1) & 483 (9) \\  
HC$_3$N \hfill $3-2,4-3$ & 27294.35 & 3607 (54) & 7.15 & 0.16 (0) & 614 (9) & & 1760 (22) & $-$3.93 & 0.39 (1) & 732 (10) \\  
HC$_3$N \hfill $3-2,2-2$ & 27296.23 & 525 (53) & 7.08 & 0.16 (2) & 89 (14) & &  246 (66) & $-$4.02 & 0.36 (11) & 94 (38) \\
HC$_5$N \hfill $11-10$& 29289.15  & 1364 (40) & 7.05 & 0.21 (1) & 305 (8) & &998 (22) & $-$4.05 & 0.37 (1) & 393 (9)\\
HC$_5$N \hfill $12-11$& 31951.77  & 1650 (23) & 7.08 & 0.20 (0) & 351 (5) & &\nodata &  \nodata & \nodata & \nodata \\
HC$_5$N \hfill $13-12$& 34614.39  & 1662 (24) & 7.13 & 0.18 (0) & 318 (5) & &\nodata &  \nodata & \nodata & \nodata \\
HC$_7$N \hfill $25-24$ & 28199.81 & 202 (8) &  7.20 & 0.14 (1) & 30 (1) & &  123 (11) & $-$3.92 & 0.29 (3) & 38 (3) \\
C$_3$S \hfill $5-4$ & 28903.69 & 573 (45) & 7.16 & 0.15 (1) & 92 (7) &  &  517 (21) & $-$4.01 & 0.33 (2) & 182 (7)\\
CH$_3$CCH \hfill $2_1-1_1$ & 34182.76 & 252 (25) & 7.13 & 0.19 (2) & 51 (5) & &\nodata &  \nodata & \nodata & \nodata \\
CH$_3$CCH \hfill $2_0-1_0$ & 34183.42 & 264 (22) & 7.15 & 0.23 (2) & 65 (5) & &\nodata &  \nodata & \nodata & \nodata \\
\enddata
\tablenotetext{a} {Transitions are specified as follows: C$_6$H$^-$, HC$_5$N, HC$_7$N \& C$_3$S: $J''-J'$; C$_6$H: $J''-J'$,\,parity,\,$F''-F'$; C$_4$H: $N''-N',J''-J',F''-F'$; HC$_3$N: $J''-J',F''-F'$; CH$_3$CCH: $J_K''-J_K'$.}
\tablecomments{Units of $T_a$ are in mK and velocities are in km\,s$^{-1}$, relative to the LSR frame. The $1\sigma$ errors on the last quoted digit(s) are given in parentheses.}
\end{deluxetable*}

\begin{deluxetable}{lrr}
\tabletypesize{\footnotesize}
\tablecaption{Observed molecular column densities\label{tab:colds}}
\tablewidth{0pt}
\tablehead{
\colhead{Species}&\colhead{L1512}&\colhead{L1251A}
}
\startdata
C$_6$H$^-$&2.0$\pm0.5\times10^{10}$ &   2.7$\pm0.7\times10^{10}$\\
C$_6$H &   4.8$\pm0.3\times10^{11}$  &  7.6$\pm0.8\times10^{11}$ \\
C$_4$H &   9.2$\pm0.6\times10^{13}$ &   1.2$\pm0.1\times10^{14}$ \\
HC$_3$N &  3.0$\pm0.4\times10^{13}$ & 2.8$\pm0.8\times10^{13}$\\
HC$_5$N &  4.9$\pm0.1\times10^{12}$  &  7.5$\pm0.2\times10^{12}$\\
HC$_7$N &  1.9$\pm0.1\times10^{12}$  &  4.7$\pm0.4\times10^{12}$\\
C$_3$S &   1.3$\pm0.1\times10^{12}$  &  2.6$\pm0.1\times10^{12}$\\
CH$_3$CCH &3.1$\pm0.3\times10^{13}$& \nodata\\ 

\enddata
\tablecomments{Units are cm$^{-2}$.}
\end{deluxetable}

The C$_6$H$^-$ and C$_6$H spectra observed in L1251A and L1512 are shown in Figure \ref{fig:spectra}. To improve the signal-to-noise ratio for L1251A, the C$_6$H$^-$ $J=10-9$ and $J=11-10$ spectra have been averaged in velocity space.

The observed spectral lines were least-squares fitted using single Gaussians, the parameters for which are given in Table \ref{tab:lines}. The five hyperfine peaks of the HC$_3$N $J=3-2$ transition are well-resolved in our spectra, from which we derived gas rotational excitation temperatures ($T_{ex}$) of $6.2\pm0.3$~K for L1251A and $8.7\pm0.7$~K for L1512 (using Equations 1 and 2 of \citealt{sav02}). These temperatures were used in the calculation of the column densities of the observed species assuming LTE and optically thin emission (see for example, Equation 2 of \citealt{lis02}). Where multiple transitions were observed for a given species, the average of the calculated column densities was taken, except for HC$_3$N, for which only the optically thin $\Delta F=0$ transitions were used. The HC$_3$N and other recorded spectra will be presented in a future article. Column densities are given in Table \ref{tab:colds}.

\section{Discussion}

Large abundances of carbon-chain-bearing species were observed in both L1251A and L1512. The measured C$_4$H and C$_6$H column densities (see Table \ref{tab:colds}) are similar to those in interstellar clouds with the highest known carbon chain abundances in the Galaxy \citep[see][]{sak08b,sak09,gup09}. The abundance of carbon chains is also highlighted in L1512 by the large CH$_3$CCH column density, which is comparable to that found in L1527 and TMC-1 \citep{sak08b}. The C$_3$S column densities in both L1251A and L1512 are somewhat less than previously observed in `carbon-chain-producing regions' by \citet{hir06}. Possible explanations for this include depletion onto dust or reduced elemental sulphur abundances. 

The derived C$_6$H$^-$ column densities of 2.7$\pm0.7\times10^{10}$~cm$^{-2}$ in L1251A and 2.0$\pm0.5\times10^{10}$~cm$^{-2}$ in L1512 are similar to those previously observed in other low-mass star-forming cores, \eg\ L1521F and L1544 (3.4$\times10^{10}$~cm$^{-2}$ and 3.1$\times10^{10}$~cm$^{-2}$, respectively; \citealt{gup09}), and L1527 (5.8$\times10^{10}$~cm$^{-2}$; \citealt{sak07}).  These values are somewhat less than those observed in the quiescent molecular clouds TMC-1 (1.2$\times10^{11}$~cm$^{-2}$; \citealt{bru07}) and Lupus-1A (6.5$\times10^{10}$~cm$^{-2}$; \citealt{sak10}). The anion-to-neutral ratios ([C$_6$H$^-]/[$C$_6$H]), however, are approximately in the middle of the previously observed distribution: $3.6\pm1.3$\% for L1251A and $4.2\pm1.4$\% for L1512, compared with 1.6\% for TMC-1, 2.1\% for Lupus-1A, 2.5\% for L1544, 4\% for L1521F and 9.3\% for L1527. \citet{sak07,sak10} hypothesised that an inverse relationship between H-atom abundance and gas density would result in greater [C$_6$H$^-]/[$C$_6$H] ratios in denser gas, which may explain the larger ratio found in L1527 where the hydrogen nucleon density $n_H$ is $\sim10^6$~cm$^{-3}$, compared to $n_H\sim10^4$~cm$^{-3}$ in TMC-1. L1251A and L1512 fit this trend -- both have $n_H\sim10^5$~cm$^{-3}$ \citep{lee10,kir05}, and their anion-to-neutral ratios are intermediate between TMC-1 and L1527. 

The detection of C$_6$H$^-$ in L1251A is the first reported interstellar anion in a protostar outside of Taurus, and thus shows that the importance of anions must be considered in future studies of the chemistry of star forming regions throughout the Galaxy.

L1251A and L1512 have similar [C$_6$H]/[C$_4$H] ratios of and 0.6\% and 0.5\%, respectively.  These are within the range of values previously measured in interstellar clouds by \citet{gup09}.

\subsection{L1251A}
\label{sec:l1251a}

The L1251A molecular cloud is located in the Cepheus Flare region of low-to-intermediate mass star formation \citep{kun08}. Our observed position is $40''$ SE of the embedded Class 0 protostar L1251A IRS3, which powers a molecular outflow \citep{lee10}. The protostar centre is located outside of the $26''$ GBT beam (see Figure \ref{fig:map}), so emission from its core does not directly influence our observations. However, depending on the radius of its outer boundary (which is likely to be up to a few times $10^4$~AU; see, \eg\ \citealt{jor02}), gas from inside the protostar envelope is probably responsible for much of the observed molecular emission. It would be of interest to observe the carbon chain (and anion) emission closer to the protostellar core, to determine whether the elevated temperatures there have any impact on the abundances of these species, as has been hypothesised by \citet{sak08b}.

Although previous observations have shown relatively large HC$_3$N and HC$_5$N abundances in various parts of the Cepheus complex, our observations constitute the largest reported C$_4$H column density and the first detection of the long-chain species C$_6$H and HC$_7$N in this region. The derived [HC$_7$N]/[HC$_5$N] ratio of 63\% is exceptionally high compared with other dense interstellar clouds; the largest value observed in the seven carbon-chain-rich clouds reported by \citet{hir06} and \citet{sak08b} is 37\% in TMC-1. However, there is some uncertainty in the excitation temperatures of these species, as they have been found to differ from each other in TMC-1 by about 1-3~K \citep[\eg][]{bel98}.  Assuming the HC$_5$N rotational temperature is 4-5~K and the HC$_7$N temperature 5-7~K, and accounting for the optical depth of the HC$_5$N line (0.3), the resulting [HC$_7$N]/[HC$_5$N] ratio in L1251A is 17-89\%.  \citet{cer86} observed an average ratio of 28\% in six cloudlets in Taurus. Thus, the ratio in L1251A is likely to be similar to or larger than in Taurus. These results show that chemical conditions in the Cepheus Flare can be highly conducive to the formation of relatively long carbon chain-bearing species.

The profiles of the spectral lines in L1251A show a small amount of asymmetry -- insufficient to justify fitting multiple Gaussians, but nevertheless indicative of a more complex cloud structure.  Although we derive a lower temperature for this cloud than L1512, the lines are broader by approximately a factor of two, which indicates the presence of significant bulk motions of the gases along the line of sight.  This may be related to turbulence arising from the L1251A IRS3 outflow, which partially intersects our observed telescope beam.

There is some evidence that the C$_6$H$^-$ lines (with $\Delta v\approx0.20$~\kms) are narrower in this source than the lines of the neutral species (with $\Delta v\approx0.30$~\kms). A similar phenomenon was also observed by \citet{sak10} in Lupus-1A, who suggested that this may be due to the anion being preferentially located in cooler, denser gas than the neutral. This suggestion seems reasonable on the basis that the rate of radiative electron attachment, and thus, the rate of anion formation is theorised to be proportional to $T^{-1/2}$ and to the electron density \citep[][]{her08}.

\subsection{L1512}

L1512 is a rather isolated molecular cloud core located in the nearby Taurus-Auriga complex \citep{ung87}. It contains a compact (${\rm FWHM}\sim10^4$~AU) submillimeter source centered around RA(2000) = 5:04:08.2, DEC(2000) = +32:43:32 \citep{dif08,kir05}, $26''$ NE of our observed position (see Figure \ref{fig:map}), which indicates the presence of a dense pre-stellar core covering our telescope beam. \citet{gup09} previously attempted to detect C$_6$H$^-$ at the \citet{ben89} position ($30''$ east of our observed position, as shown in Figure \ref{fig:map}), and they derived an upper limit of 4.8$\times10^{10}$~cm$^{-2}$, which is consistent with our observed column density (see Table \ref{tab:colds}).  They also derived C$_4$H and C$_6$H column densities of 9$\pm3\times10^{13}$~cm$^{-2}$ and 5$\pm2\times10^{11}$~cm$^{-2}$, respectively, which closely match our observed values. \citet{hir09} measured an HC$_3$N column density of 2.5$\times10^{13}$~cm$^{-2}$, $10''$ west of our L1512 position, which is in reasonably good agreement with our observed value. The similarities of our observed column densities with those found nearby indicate that polyyne and cyanopolyyne abundances may be less variable than shown by the HC$_3$N map in Figure \ref{fig:map}, over spatial scales of $\sim30''$ (which corresponds to 0.02~pc at the distance of L1512).  This implies that the strong, compact peak shown in the HC$_3$N map may arise at least partly as a result of enhanced gas densities and/or temperatures in that region, which (in the case of sub-critical line excitation), would cause increased HC$_3$N $J=10-9$ emission as a result of increased excitation of the relatively high energy (24~K) $J=10$ rotational level.

\subsection{Comparison and Carbon Chemistry}

Located in two distinctly separate parts of the sky (L1251A is $\sim330$~pc distant in Cepheus, whereas L1512 is $\sim140$~pc distant in Auriga), it is surprising how chemically similar these two clouds are. Apart from HC$_7$N, all of the observed column densities are within a factor of two of each other (see Table \ref{tab:colds}). This also applies to NH$_3$ \citep{ben89} and N$_2$H$^+$ \citep{cas02}.  The large anion, polyyne and cyanopolyyne abundances are indicative of an active carbon chemistry, comparable to that observed in other carbon-chain-rich regions of the Galaxy (see \eg\ \citealt{hir06} and \citealt{hir09}).

Large abundances of unsaturated carbon chains are theorised to occur either in the relatively early stages of dark cloud evolution when carbon is abundant in reactive form (see for example, \citealt{her89}), or later on during the `freeze-out peak' \citep{bro90,ruf97}. In the latter scenario, the lighter, more reactive elements such as atomic oxygen are theorised to freeze out onto the dust grains over time. The loss of oxygen from the gas-phase results in reduced destruction rates for the carbon-chain-bearing species and their associated reagents, giving rise to elevated abundances of these species later on in a dense cloud's evolution.

Over time, two of the primary destructive reactants for organic anions -- atomic hydrogen and oxygen -- are lost from the gas phase through conversion on dust grain surfaces to H$_2$ and H$_2$O, respectively \citep{gol05,ber00}. This may explain why the C$_6$H$^-$ anion-to-neutral ratios are greater in denser clouds (including L1521A, L1512 and L1527), than in the quiescent, lower-density clouds TMC-1 and Lupus-1A.  Oxygen and hydrogen react quickly with C$_6$H$^-$ \citep{eic07}, so the combined effects of the depletion of O and H in these relatively more dense environments may be responsible for the large observed anion-to-neutral ratios in protostars and pre-stellar cores.

Alternatively, elevated carbon chain abundances could arise as a consequence of selective methane sublimation from dust grain surfaces, as originally investigated by \citet{mil84} and \citet{bro91}. It is hypothesised that as a protostellar object contracts and begins to heat the surrounding gas, sublimated methane reacts with C$^+$ to form hydrocarbon ions, which gives rise to a so-called warm carbon chain chemistry (WCCC) via subsequent ion-molecule reactions (see \citealt{sak08b} and \citealt{has08}).  Given the close proximity of our observed L1251A position to the protostar IRS3, the WCCC theory may be applicable as a possible explanation for the large hydrocarbon abundances present there.

The location of our L1512 anion detection coincides with the region of CO depletion identified by \citet{buc06}. This is also the case for the anion and carbon chain detections in L1521F and L1544 made by \citet{gup09}. Thus, the `freeze-out peak' chemistry may be the preferred explanation for the large anion and carbon chain abundances observed in L1512 and similar pre-stellar cores. The presence of hydrocarbon anions is likely to be partly responsible for the large (neutral) polyyne and cyanopolyyne abundances in L1251A and L1512 \citep[see][]{wal09}.

\section{Conclusion}

We have detected the carbon chain anion C$_6$H$^-$ for the first time in L1251A and L1512. Large column densities of the neutral species C$_4$H, C$_6$H, HC$_5$N, HC$_7$N and C$_3$S were also detected. Combined with the five previous interstellar anion detections by \citet{mcc06}, \citet{gup09} and \citet{sak07,sak10}, these results show that anions tend to be present in detectable quantities in regions where carbon chains are highly abundant, especially surrounding embedded protostars and pre-stellar cores. The anion-to-neutral ratio [C$_6$H$^-$]/[C$_6$H] is consistently on the order of a few percent and shows an apparent positive correlation with density. This implies that a relatively uniform and simple physical/chemical mechanism is responsible for regulating molecular anion abundances in the different interstellar clouds observed so far.

The large polyyne and cyanopolyyne abundances measured in L1251A show that carbon chain production processes are rapid in this part of the Cepheus Flare molecular cloud complex, resulting in similar abundances to those found in the Taurus Molecular Cloud. The observation of abundant C$_6$H$^-$ in L1251A constitutes the first detection of anions in a protostellar envelope outside of the Taurus-Auriga complex, and indicates that anions are likely to be widespread throughout Galactic star-forming regions where carbon chains are present.

The technique of using HC$_3$N and C$_4$H as proxies for the detection of less abundant, larger carbon chains and anions shows strong promise as a means for obtaining further detections of these molecules in low-mass star-forming regions in the future.

\acknowledgments
This research was supported by the NASA Exobiology Program and the Goddard Center for Astrobiology. Astrophysics at QUB is supported by a grant from STFC.


\begin{thebibliography}{}
\bibitem[Ag{\'u}ndez et al.(2008)]{agu08}Ag{\'u}ndez, M., Fonfr{\'i}a, J. P., Cernicharo, J., Pardo, J. R., Gu{\'e}lin, M. 2008, A\&A, 479, 493
\bibitem[Ag{\'u}ndez et al.(2010)]{agu10}Ag{\'u}ndez, M., Cernicharo, J., Gu{\'e}lin, M. et al.. 2010, A\&A, 517, L2
\bibitem[Benson \& Myers(1989)]{ben89}Benson, P. J., Myers, P. C. 1989, ApJS, 71, 89
\bibitem[Bell et al.(1998)]{bel98}Bell, M. B., Watson, J. K. G., Feldman, P. A., Travers, M. J. 1998, ApJ, 508, 286
\bibitem[Bergin et al.(2000)]{ber00}Bergin, E. A., Melnick, G. J., Stauffer, J. R. et al.. 2000, ApJ, 539, 129
\bibitem[Bergin et al.(2007)]{ber07}Bergin, E. A. \& Tafalla, M. 2007, ARA\&A, 45, 339
\bibitem[Brown \& Charnley(1990)]{bro90}Brown, P. D., Charnley, S. B. 1990, MNRAS, 244, 432
\bibitem[Brown \& Charnley(1991)]{bro91}Brown, P. D., Charnley, S. B. 1991, MNRAS, 249, 69
\bibitem[Br{\"u}nken et al.(2007)]{bru07}Br{\"u}nken, S., Gupta, H., Gottlieb, C. A., McCarthy, M. C., Thaddeus, P. 2007, 664, L43
\bibitem[Buckle et al.(2006)]{buc06}Buckle, J. V., Rodgers, S. D., Wirstrom, E. S., Charnley, S. B., Markwick-Kemper, A. J., Butner, H. M., Takakuwa, S. 2006, Faraday Discussions, 133, 63
\bibitem[Caselli et al.(2002)]{cas02}Caselli, P., Benson, P. J., Myers, P. C., Tafalla, M. 2002, ApJ, 572, 238 
\bibitem[Cernicharo et al.(1986)]{cer86}Cernicharo, J., Bachiller, R., Duvert, G. 1986, A\&A, 160, 181
\bibitem[Cordiner \& Sarre(2007)]{cor07}Cordiner, M. A. \& Sarre, P. J. 2007, A\&A, 472, 537
\bibitem[Cordiner et al.(2008)]{cor08}Cordiner, M. A., Millar, T. J., Walsh, C., Herbst, E., Lis, D. C., Bell, T. A., Roueff, E. 2008, IAUS, 251, 157
\bibitem[Cordiner \& Millar(2009)]{cor09} Cordiner, M. A. \& Millar, T. J. 2009, ApJ, 697, 68
\bibitem[Cernicharo et al.(2007)]{cer07}Cernicharo, J., Gu{\'e}lin, M., Ag{\'u}ndez, M., Kawaguchi, K., McCarthy, M., Thaddeus, P. 2007, A\&A, 467, 37
\bibitem[Di Francesco et al.(2008)]{dif08}Di Francesco, J., Johnstone, D., Kirk, H., MacKenzie, T., Ledwosinska, E. 2008, ApJS, 175, 227
\bibitem[Eichelberger et al.(2007)]{eic07}Eichelberger, B., Snow, T. P., Barckholtz, C., Bierbaum, V. M. 2007, ApJ, 667, 1283
\bibitem[Federman et al.(1990)]{fed90}Federman, S. R., Huntress, W. T., Jr., Prasad, S. S., 1990, ApJ, 354, 504
\bibitem[Gupta et al.(2009)]{gup09}Gupta, H., Gottlieb, C. A., McCarthy, M. C., Thaddeus, P. 2009, ApJ 691, 1494
\bibitem[Goldsmith \& Li(2005)]{gol05} Goldsmith, P. F., Li, D. 2005, ApJ, 622, 938
\bibitem[Hassel et al.(2008)]{has08}Hassel, G. E., Herbst, E., Garrod, R. T. 2008, ApJ, 681, 1385
\bibitem[Harada \& Herbst (2008)]{har08}Harada, N., Herbst, E. 2008, ApJ, 685, 272
\bibitem[Herbst(1981)]{her81}Herbst, E. 1981, Nature, 289, 656
\bibitem[Herbst \& Leung(1989)]{her89}Herbst, E., Leung, C. M. 1989, ApJSS, 69, 271
\bibitem[Herbst \& Osamura(2008)]{her08}Herbst, E., Osamura, Y. 2008, ApJ, 679, 1670
\bibitem[Hirota \& Yamamoto(2006)]{hir06} Hirota, T., Yamamoto, S. 2006, ApJ, 646, 258
\bibitem[Hirota et al.(2009)]{hir09} Hirota, T., Ohishi, M., Yamamoto, S. 2009, ApJ, 699, 585
\bibitem[J{\o}rgensen et al.(2002)]{jor02}J{\o}rgensen, J.~K., Sch{\"o}ier, F.~L., van Dishoeck, E.~F. 2002, ApJ, 389, 908
\bibitem[Kirk et al.(2005)]{kir05}Kirk, J. M., Ward-Thompson, D., Andr{\'e}, P. 2005, MNRAS, 360, 1506
\bibitem[Kun et al.(2008)]{kun08}Kun, M., Kiss, Z. T., Balog, Z. 2008, In Book: Handbook of Star Forming Regions, Volume I, 136
\bibitem[Little et al.(1978)]{lit78}Little, L. T., Riley, P. W., MacDonald, G. H., Matheson, D. N. 1978, MNRAS, 183, 805
\bibitem[Lee et al.(2010)]{lee10}Lee, J., Lee, H., Shinn, J., Dunham, M., Kim, I., Kim, C. H., Bourke, T. L., Evans, N. J., Choi, Y. 2010, ApJ, 709, 74
\bibitem[Lis et al.(2002)]{lis02}Lis, D. C., Roueff, E., Gerin, M., Phillips, T. G., Coudert, L. H., van der Tak, F. F. S., Schilke, P. 2002, ApJ 571, L55
\bibitem[McCarthy et al.(2006)]{mcc06}McCarthy, M. C., Gottlieb, C. A., 
  Gupta, H. C., \& Thaddeus, P. 2006, \apjl, 652, L141
\bibitem[Millar \& Freeman(1984)]{mil84}Millar, T. J., Freeman, A. 1984, MNRAS, 207, 405
\bibitem[Millar \& Herbst(1994)]{mil94}Millar, T. J., Herbst, E. 1994, A\&A, 288, 561
\bibitem[Millar et al.(2007)]{mil07}Millar, T. J., Walsh, C., Cordiner, M. A., N{\'i} Chuim{\'i}n, R., Herbst, E. 2007, {ApJL}, 662, L87
  R. P. A. 2000, \mnras, 316, 195
\bibitem[Remijan et al.(2007)]{rem07}Remijan, A. J., Hollis, J. M., Lovas, F. J., Cordiner, M. A., Millar, T. J., Markwick-Kemper, A. J., Jewell, P. R. 2007, {ApJL}, 664, L47
\bibitem[Ruffle et al.(1997)]{ruf97}Ruffle, D. P., Hartquist, T. W., Taylor, S. D., Williams, D. A. 1997, MNRAS, 291, 235
\bibitem[Sakai et al.(2007)]{sak07}Sakai, N., Sakai, T., Osamura, Y., Yamamoto, S. 2007, ApJL, 667, L71
\bibitem[Sakai et al.(2008a)]{sak08a}Sakai, N., Sakai, T., Yamamoto, S. 2008, ApJL, 673, L71
\bibitem[Sakai et al.(2008b)]{sak08b}Sakai, N., Sakai, T., Hirota, T., Yamamoto, S. 2008, ApJ, 672, 371
\bibitem[Sakai et al.(2009)]{sak09}Sakai, N., Sakai, T., Hirota, T., Yamamoto, S. 2009, ApJ, 702, 1025
\bibitem[Sakai et al.(2010)]{sak10}Sakai, N., Shiino, T., Hirota, T., Sakai, T., Yamamoto, S. 2010, ApJ, 718, L49
\bibitem[Sarre(1980)]{sar80}Sarre, P. J. 1980, J. Chim. Phys., 77, 769
\bibitem[Savage et al.(2002)]{sav02}Savage, C., Apponi, A. J., Ziurys, L. M., Wyckoff, S. 2002, ApJ, 578, 211
\bibitem[Ungerechts \& Thaddeus(1987)]{ung87}Ungerechts, H., Thaddeus, P 1987, ApJSS, 63, 645
\bibitem[Walsh et al.(2009)]{wal09}Walsh, C., Harada, N., Herbst, E., Millar, T. J. 2009, ApJ, 700, 725
\bibitem[Woodin et al.(1980)]{wood80} Woodin, R., Foster, M. S., \& Beauchamp, J. L. 1980, J. Chem. Phys., 72, 4223
\end{thebibliography}
\end{document}